\documentstyle[12pt]{article}
\def\beqa{\begin{eqnarray}}
\def\eeqa{\end{eqnarray}}
\def\beq{\begin{equation}}
\def\eeq{\end{equation}}

\def\half{\frac{1}{2}}

\def\hm{h^{-1}{\rm~Mpc}}
\def\vr{\mbox{\bf r}}

\def\vx{\mbox{\bf x}}
\def\vy{\mbox{\bf y}}
\def\cz{\chi_0}

\let\lam=\lambda  
\let\eps=\varepsilon
\let\gam=\gamma

\newcommand{\mincir}{\raise -2.truept\hbox{\rlap{\hbox{$\sim$}}\raise5.truept
\hbox{$<$}\ }}
\newcommand{\magcir}{\raise -2.truept\hbox{\rlap{\hbox{$\sim$}}\raise5.truept
\hbox{$>$}\ }}
\newcommand{\minmag}{\raise-2.truept\hbox{\rlap{\hbox{$<$}}
\raise 6.truept\hbox
{$>$}\ }}

\def\etal{{\it et al.}\ }

\def\rprl{ Phys. Rev. Lett. }

\def\rapj{ Ap. J. }

\def\rmnras{ Mon. Not. R. Ast. Soc. }

\begin{document}
%\draft
\begin{titlepage}
\thispagestyle{\empty}

%\null
%\vspace{.3in}
%\begin{flushright}
%Fermilab
%\end{flushright}
\vspace{.2in}

\vspace{.6in}
\centerline{\Large\bf Non-Gaussian  Likelihood Function }
\vspace{.4in}
\centerline{\bf Luca Amendola}
\vspace{.1in}
\centerline{Osservatorio Astronomico di Roma
\footnote{e-mail amendola@oarhp1.rm.astro.it}}
\centerline{Viale del Parco Mellini, 84,  Rome 00136 - Italy}
\vspace{.6in}
%\address{}
%\date{\today}

\begin{abstract}
\baselineskip 14pt
We generalize the maximum likelihood method to non-Gaussian
distribution functions by means of the multivariate  Edgeworth expansion.
We stress the potential interest of this technique in all those
cosmological problems in which the determination
of a non-Gaussian signature is relevant, e.g.
in the analysis of
 large scale structure and cosmic microwave background.
A first important result is that the asymptotic
confidence limits on the
parameters are systematically widened  when the higher
order correlation functions are positive, with respect to
the Gaussian case.
\end{abstract}
\vspace{.8in}
{\bf Subject headings}: cosmology: large-scale structure of the Universe;
galaxies: clustering;
methods: statistics
\end{titlepage}

%\pacs{98.80.Bp,Dr}
\section{Introduction}
\baselineskip 14pt

\label{sec:intro}
Modern large scale astronomy is, to a large extent, the science of
non-Gaussian random fields. One of the keys to understanding the
formation and evolution of structure in the Universe resides infact
in the statistical properties of the matter field. Rival
theories of structure formation predicts different statistical features,
both in the present Universe and in the primordial fluctuations
encoded
in the microwave background. To the scope of quantifying the
statistical feature of the matter clustering, several techniques have
been proposed. One of these, which is increasingly popular in
astrophysics, is the estimation of parameters via the maximum likelihood
method. For instance, the maximum
likelihood method is currently widely employed in the
analysis of cosmic microwave background
(CMB) experiments, large scale surveys and cosmic velocity fields.
Once a suitable likelihood function (LF) has been constructed,
one estimates the best parameters  simply by finding the maximum
of the LF with respect to those parameters.
The parameter estimates, say $\hat\alpha_i(x_i)$ (the hat is to distinguish
between the estimates and the theoretically expected parameters),
 are then functions of the data $x_i$, i.e. of the
random variables, and are therefore random variables themselves.
If one is able to determine the distribution function $P(\hat\alpha_i)$
of the estimators, the confidence region (CR) of the
parameters can be found as the value of the parameters for
which the integral of $P(\hat\alpha_i)$
 falls below a predetermined level.

There are however two problems with this approach.
One is that usually
we don't know how the data are distributed, and the
usual  Gaussian approximation
may be very poor. This is the case, for instance,
 in  large scale structure, where we already know that
the density fluctuations are not Gaussian, even on fairly large scales.
The second problem is that, even if we know perfectly well
the data distribution, is often not trivial to find an analytical
expression for the distribution $P(\hat\alpha_i)$
of the parameter estimators $\hat\alpha_i$. Aside the simplest case
in which one only needs the raw sample variance or the sample mean of
normal variates (or closely related quantities),
one has invariably to resort to very time-consuming MonteCarlo
methods.
While this second problem can be always overcome by numerical
methods, the first difficulty remains, unless one takes into consideration
specifically designed non-Gaussian models,
 and for each of these determines the
confidence regions for the relevant parameters.
Other than being too model-dependent,
in the current astrophysical applications this procedure
is in many cases prohibitively slow.
It is then of interest to examine alternatives able to retain
the useful features of the  likelihood
method while allowing more freedom in exploring
different non-Gaussian (non-G, for shortness) distributions.

In this work we propose a perturbative method to estimate theoretical
parameters when the higher-order multivariate moments (or $n$-points
correlation functions) are non-vanishing,
via an expansion around a Gaussian LF, the
multivariate Edgeworth expansion (MEE).
As long as the perturbative approach does
not break down, i.e. as long as the departure from Gaussianity is mild,
the MEE gives an answer to the first problem, the distribution function
of the data, because it allows
 arbitrary values of the higher-order correlation functions.
 Then we still are left with the second problem: how do we determine
in the general case the
distribution function for our parameter estimators,
 necessary to produce the confidence regions?
A first simple possibility is to approximate the $P(\hat\alpha_i)$
around its peak, previously determined by maximization of the LF,
by a Gaussian distribution, multivariate in the parameter space.
This allows to determine an approximate covariance matrix, whose
eigenvalues give the principal axis of the parameter CR (see e.g.
Kendall, Stuart \& Ord 1987). Notice that this is
equivalent to assume the {\it parameter estimators},
which are functions of all the dataset,
 as Gaussian distributed, but makes no
assumptions on  the {\it data} themselves.
 When the
number of data is large, this procedure can be justified by
the central limit theorem (which however cannot guarantee the
asymptotic Gaussianity in the general case). This first possibility,
along with its limitations,
is discussed in Sect. 3.
To overcome  the limits relative to the Gaussian
approximation of the estimator distribution, and exploiting the
analytic properties of the MEE, we adopt in Sect. 4 and Sect. 5
a second, exact, way to determine the CR for some
of the relevant parameters.
This is a non-Gaussian generalization of the
$\chi^2$ technique: instead of finding the CR by integration over
the unknown distribution function
of the sample estimators, we determine the CR by integrating the LF
over the possible outcomes of an experiment.
As in the usual $\chi^2$ method, the acceptable values of a parameter
will be all those for which  the data lie not too distant from the
predicted values, the ``distance'' being measured by the quantity
$\chi^2_0=x_i\lam^{ij} x_j$, where $x_i$ are the actual data and
$\lam^{ij}$ is the inverse of the correlation matrix. The CR
will in general depend on all the higher-order correlation
functions included in the MEE, as it will be shown in Sect. 4.
 The formalism is then best suited
to answer the question: how our results (i.e., best
estimates and  CR) change when the higher-order
moments of the data distribution are not set to zero?
If we have any reason to believe in some particular values for
the non-G moments, then we can plot the CR
for our parameters given those higher-order moments, and clearly the
regions will be different for any set of higher-order moments.
The relevance of examining how the confidence regions vary
with respect to the non-G parameters is clear. Suppose
in fact that
two  experiments, assuming Gaussianity,
 produce two non-overlapping CR
 for the same parameter, say the overall normalization of the
correlation function. In
general, the CR will be different in the
non-Gaussian case, and it may happen that the two experiments are
infact compatible when some level of non-Gaussianity is assumed.
As we will show, in most cases the CR widens for positive
higher-order moments, so that two non-overlapping results can
be brought to agreement, {\it provided some amount of
non-Gaussianity is allowed}.
 Other positive features of our
formalism are that it exploits the full set of data, that it can
be extended to higher and higher order moments, and that it is
fully analytical.

To the order to which we limit ourselves here,
we will be able to
 estimate the first non-G correlation function, i.e. the
third-order moments.
This estimate will share the good and less good properties of
likelihood estimators: they are consistent estimators, but
only asymptotically  (i.e., for large samples)
unbiassed, as we will show in Sect. 3.
For the fourth-order cumulant there is not an
estimator at all, since the LF is linear in it.
We decided to keep track of it anyway, because
it is still interesting  to use the fourth-order cumulant as an
external parameter,
 and see how our results change with different assumptions on it.

In principle one can include in the analysis all the set of
higher-order moments considered relevant to the problem, but here we
will limit ourselves to the first two higher-order terms, the
3- and 4-points correlation functions. For most purposes, this is
the best we can do for comparing different models with
observations, since current data
do not permit accurate analysis of correlation functions of order
higher than the fourth one.

 Let us remark that we call here likelihood
function the probability
distribution function $f(x_i,\alpha_i)$
of the data $x_i$ (our random variables) defined in a
sample space $S$,
 given some theoretical
parameter $\alpha_i$, which can be
thought to lie in the parameter space
$P$. Essentially, for any point in the parameter space,
i.e. for any  distribution function,
we will integrate the LF
over the sample space $S$, i.e. over all
the possible outcomes of the
experiment, to determine how likely or unlikely is the
possibility that the actual data set has arisen from such a
 parametric choice. For a discussion of the
advantage and disadvantage of this approach with respect to the
alternative Bayesan one, in which the integration occurs over
the  space of the {\it theoretical} parameters
(as opposed to the {\it sample estimators} of the
theoretical parameters considered in the
frequentist approach), we refer to standard textbooks like
Kendall, Stuart \& Ord (1987).

Beside presenting the basic formalism of the non-Gaussian  LF,
we discuss briefly in Sect. 6
its application to large scale structure and to the CMB.
In the first case the non-Gaussian nature of the galaxy distribution is
a well-established fact, so that the use of a non-G LF is
certainly required. In the case of CMB, the current set of data is
still not accurate enough to assess the issue. The estimate of a confidence
region in non-G models  is however crucial in view of the
discrimination among different theories of structure formation.

\section{Formalism}
Let $d^i$ be a set of experimental data, $i=1,..N$,
and let us form the variables $x^i=d^i-t^i$, where $t^i$ are
the theoretical expected values for the measured quantities.
To fix the ideas, one can think of $d^i$ as the temperature fluctuation
in the $i$-th pixel in a CMB experiment, or as the number of galaxies
in a given volume of the Universe.
Let $c^{ij}$ be  the correlation matrix
\beq\label{cm}
c^{ij}=<x^ix^j>\,,
\eeq
and let us introduce the higher-order cumulant matrices
(or $n$-point correlation functions)
\beqa
k^{ijk}&=&<x^i x^j x^k>\qquad{\rm~(skewness~matrix)},\\
k^{ijkl}&=&<x^i x^j x^k x^l>-c^{ij}c^{kl}-c^{ik}c^{jl}-c^{il}c^{jk}
\qquad {\rm~(kurtosis~matrix)}
\eeqa
(we will sometimes use the words ``skewness'' and ``kurtosis''
to refer to the 3- and 4-point correlation functions, respectively,
or to their overall amplitude;
in the statistical literature, the definition of
skewness is actually, in our notation, $\gam_1=k^{iii}/(c^{ii})^{3/2}$,
and for the kurtosis $\gam_2=k^{iiii}/(c^{ii})^2$).
The correlation matrices depend in general both on a
number of theoretical parameters $\alpha_j$, $j=1,..P$
(that we leave for the moment unspecified) and on the experimental errors.
In most cases, we can assume the experimental errors
to be
Gaussian distributed (or even uncorrelated) so that they can be
completely characterized by the correlation matrix $e^{ij}$,
which is simply to be
added in quadrature to  the 2-point
correlation function.
It is useful to define then the matrix
\beq
\lam_{ij}=(c^{ij}+e^{ij})^{-1}\,.
\eeq
The problem of estimating the parameters $\alpha_j$
is solved by maximizing, with respect to the
parameters, the likelihood
function
\beq
L=f(\vx)\,,
\eeq
where $f(\vx)$ is the  multivariate probability
distribution function (PDF) for
 the random variables $x_i$.
Clearly, knowing the LF one can, at least in principle,
determine also the parameter CR, as will be discussed
in the next sections.
The main difficulty to this approach, however,
 is that we do not know, in general, the
exact form for the PDF $f(\vx)$. The usual
simplifying assumption is then
that $f(\vx)$ is a multivariate Gaussian distribution
\beq
L_g=f(\vx)=G(\vx,\lam)\equiv
(2\pi)^{-N/2} |\lam|^{1/2} \exp(-\half x^i \lam_{ij} x^j)\,.
\eeq
where $|\lam|={\rm det}(\lam_{ij})$.
This is usually assumed, for instance, in analysing the CMB
fluctuation maps and  the cosmic velocity fields.
A straightforward way to
generalize the LF so as to include the higher-order correlation functions,
which embody the non-Gaussian properties of the data,
is provided by the
multivariate Edgeworth expansion (MEE).
An unknown PDF $f(\vx)$ can indeed be expanded around a multivariate
Gaussian $G(x,\lam)$  according to the formula (Chambers 1967;
McCullagh 1984; Kendall, Stuart  \& Ord 1987)
\beq\label{mee}
f(\vx)=G(\vx,\lam)[1+{1\over 6}k^{ijk}h_{ijk}(\vx,\lam)
+{1\over 24 }k^{ijkl}h_{ijkl}(\vx,\lam)
+{1\over 72 }k^{ijk}k^{lmn}h_{i..n}(\vx,\lam)+...]\,,
\eeq
where $h_{ij..}$ are Hermite tensors,  the multivariate
generalizations of the Hermite polynomial. If there are $r$ subscripts,
the Hermite tensor $h_{ij..}$ is said to be of order $r$, and is given by
\beq\label{defpol}
h_{ij...}=(-1)^r G^{-1}(\vx,\lam) \partial_{ij...} G(\vx,\lam)\,,
\eeq
where $\partial_{ij...}=(\partial/\partial x_i)(\partial/\partial x_j)...$.
The Hermite polynomials
are located on the main diagonal of the Hermite tensors,
when $\lam_{ij}=\delta_{ij}$.
Notice that the function $f(\vx)$ is normalized to unity, since
the integrals of all the higher order
 terms from minus to plus infinity vanish.
It can be shown that the MEE gives a good approximation to any
distribution function provided that all the moments are defined and that
the higher order correlation functions do not dominate over the
Gaussian term. In other words, the MEE can be applied only in the limit
of mild non-Gaussianity.
More accurately, the approximation is good, in the sense that
the error one makes in the truncation is smaller than the
 terms included, if the cumulants obey the same
order-of-magnitude scaling of a standardized mean (Chambers 1967).
This condition is satisfied, for instance, by the cumulants
of the galaxy clustering in the scaling regime, which explains
why the (univariate)
Edgeworth expansion well approximates the probability
distribution of the large scale density field (Juszkiewicz \etal 1994,
Kofman \& Bernardeau 1994).
The same expansion has been also applied to the statistics of
pencil-beam surveys, in which the
one-dimensional power spectrum coefficients
 can be written as a genuine standardized mean (Amendola 1994).
Finally, it has also been used to go beyond the Gaussian approximation
in calculating the
topological genus of weakly non-Gaussian fields (Matsubara 1994).
Let us also note that the MEE lends itself to a further
generalization: if the experimental
errors are {\it not} Gaussian distributed, then
the expansion for the data given the error correlation functions
$e^{ij..}$ is the same as in Eq. (\ref{mee}), but with
the new cumulants $K^{ij..}=k^{ij..}+e^{ij..}$. In fact, let $x_t^i$
be the theoretical values whose measure is given by
 the data $d^i$,
and let $\xi^i=d^i-x_t^i$ be the experimental error.
The theoretical values are random variables in the sense that
the theory usually predicts only their distribution, not their
definite value. For instance, once the monopole is subtracted, the
standard cosmological models predict   CMB fluctuations
Gaussian distributed with zero mean, $t^i=0$. Then we
are concerned with the distribution of $x^i=
x_t^i+(d^i-x_t^i)=x_t^i+\xi^i$,
the sum of the theoretical values $x_t^i$ and of the experimental
errors $\xi^i$, both of which are
random variables.
 If the two are independent, the general theorems
on the random variables ensure that {\it the cumulants cumulate},
i.e. that the cumulants of $x^i$ are the sum of the ones of
$x_t^i$ and of $\xi^i$.

Two properties are of great help in dealing with the
MEE. The first is that $k^{ijk...}$ and $h_{ijk...}$ are
contra- and co-variant tensors, respectively, with respect to
linear transformations of the variables $x_i$. It follows then
that  $f(\vx)d\vx$  is totally invariant  with respect to the linear
transformations which leave invariant the
quadratic form $\chi^2=x^i\lam_{ij} x^j$. This
property is very useful, because we can always
diagonalize the quadratic form by choosing a linear combination
 $y^j=A^j_i x^i$ such that $\chi^2=x^i\lam_{ij} x^j= y^i \delta_{ij} y^j$.
The MEE in the new variables $y^i$ remains formally the same as in
Eq. (\ref{mee}), with $x\to y$ and
$\lam\to \delta$, but now
$G(\vy,\delta)$ factorizes, and all the calculations are simplified.
Notice that even if the new variables are uncorrelated, they are not
statistically independent, since they are not (in general)
Gaussian variates. The higher-order matrices are then not diagonalized.
In the following we will often assume that the variable transformation
has been already performed, so that we will write $y$ and $\delta$
instead of $x$
and $\lam$,
leaving all the other symbols unchanged.
The second useful property is that the MEE is {\it analytically
integrable} if the integration
region is bounded by $\chi^2=const$.
This property will be exploited in Sect. 4.

\section{Best estimates and asymptotic confidence regions}
The likelihood estimates for the parameters are to be obtained by
maximizing Eq. (\ref{mee}) with respect to the parameters.
To illustrate some interesting points, let us put ourselves in the
 simplest case, in which all data are independent, and we only
need to estimate the parameters $\sigma$ and $k_3$ entering the
2- and 3-point correlation function as overall amplitudes:
\beq\label{simple}
c_{ij}=\sigma^2 \delta_{ij},\qquad
k_{ijk}=k_3 \delta_{ij}\delta_{jk}\,.
\eeq
Because we are in such a simplified case, we will recover several
well-known formulae of sampling statistics, like the variance of
the standard deviation $\sigma$ and of $k_3$. It is important to
bear in mind, however, that the MEE is much more general than we are
assuming in this section, since it can allow
 for full correlations among data,
for experimental errors, and for non-linear parametric dependence.

 For
simplicity, we also assume  that the sample kurtosis is negligible.
Because of this, we can
put the fourth order sample cumulant of the dataset to zero
(see, e.g., Kendall, Stuart \& Ord 1987):
\beq
\hat k_4={N\over (N-1)(N-2)(N-3)}[(N+1)\sum_i x_i^4-
3(N-1)(\sum_i x_i^2)^2]=0,
\eeq
so that we have, for large $N$,
\beq\label{need}
\sum_i x_i^4= 3(\sum_i x_i^2)^2\,.
\eeq
We show here that the maximum likelihood estimators for
the variance and for the
skewness in the case of independent data and
for $N\to \infty$ reduce to
the usual sample quantities
\beqa\label{mlvar}
\hat\sigma^2&=&\sum_i x_i^2/(N-1)\,,\\
\label{sample}
\hat k_3&=&{N\over (N-1)(N-2)}\sum_i x_i^3\,.
\eeqa
We will assume also that the average has been subtracted from the data, i.e.
that $\sum_i x_i=0$. This actually reduces the degrees of freedom, but
in the limit of large $N$ we can safely ignore this problem.
If the distribution function of $x_i$ is
approximated in the limit of small
$k_3$ by the univariate Edgeworth expansion
\beq\label{uniee}
f_i=G(x_i,\sigma)[1+k_3 h_{3i}/6+k_3^2 h_{6i}/72]\,,
\eeq
where $G(x_i,\sigma)$ is a Gaussian function,
then the multivariate distribution function for the dataset is
\beq
L(\vx)=\prod_i f_i\,.
\eeq
[In the notation of Eq. (\ref{defpol}), $h_{3i}=h_{iii}$ and
$h_{6i}=h_{i..i}$.] By the definition in (\ref{defpol}) we have
\beqa\label{h3h6}
h_{6i}&=&\sigma^{-12}[x_i^6-15 \sigma^2 x_i^4+45 \sigma^4 x_i^2
-15\sigma^6]\,,\nonumber\\
h_{3i}&=&\sigma^{-6}[x_i^3-3\sigma^2 x_i]\,.
\eeqa
Let us pause to evaluate the order-of-magnitude of the
non-G corrections in the univariate Edgeworth expansion (\ref{uniee}).
Assuming $x_i\sim \sigma$, the first correction term is of the
order of $ \gam_1\equiv k_3/\sigma^3$, which
is the dimensionless definition of skewness. The general
rough requirement
for the truncated Edgeworth expansion is then that $\gam_1\ll 1$.
This condition will be encountered several times throughout this work.
The maximum likelihood estimators for $\sigma$
and $k_3$ are then the values $\hat\sigma,\hat k_3$
 which
maximize $L$, or, equivalently, its logarithm $\log L$.
We have then the equations
\beq\label{bestsig}
{d\log L\over d\sigma}=\sum_i {d\log f_i\over d\sigma}=
\sum_i\left\{
{1\over\sigma^{3}} [x_i^2-\sigma^{2}]+{k_3\over
\sigma^{7}} [2 \sigma^2 x_i-x_i^3]
+{k_3^2\over 4 \sigma^{11}}[5\sigma^4-16\sigma^2 x_i^2+5x_i^4]
 \right\}=0\,,
\eeq
and
\beq\label{bests3}
{d\log L\over dk_3}=\sum_i {d\log f_i\over dk_3}=
\sum_i\left\{ \left[{h_{3i}\over 6}+{k_3\over 36} h_{6i}\right]
/\left[1+{k_3 h_{3i}\over 6}+{k_3^2\over 72} h_{6i}\right] \right\}=0\,.
\eeq
To first order in $k_3$, the latter equation gives
\beq\label{der}
\sum_i \left[{h_{3i}\over 6}+{k_3\over 36} h_{6i}-
{k_3\over 36} h_{3i}^2\right]=0\,,
\eeq
so that our estimator is
\beq\label{k3estim}
\hat k_3=-{6\sum_i h_{3i}\over \sum_i [ h_{6i}-
 h_{3i}^2]}\,.
\eeq
Suppose now that the solution for $\sigma$ of Eqs. (\ref{bestsig}) and
(\ref{bests3}) is the usual variance estimator (\ref{mlvar}),
with $N\approx N-1$.
Then we can observe that, from (\ref{h3h6}),
\beq\label{mlh6}
\sum h_{6i}=\sum h_{3i}^2-
3\sigma^{-10}(3\sum x_i^4-12\sigma^2\sum x_i^2
+5\sum \sigma^4)=\sum h_{3i}^2-6N\sigma^{-6}\,,
\eeq
where in the last step we used Eq. (\ref{need}) and Eq. (\ref{mlvar}).
Inserting (\ref{mlh6}) in (\ref{k3estim}) we obtain finally
(assuming that $\sum_i x_i=0$)
\beq\label{k3est}
\hat k_3={\sum_i h_{3i}\over N\sigma^{-6}}=
{\sum_i x_i^3\over N}\,,
\eeq
which coincides with (\ref{sample}) for large $N$. Finally,
going back to Eq. (\ref{bestsig}), and inserting $k_3=\hat k_3$
we recover the sample variance $\hat\sigma^2=\sum_i x_i^2/N$,
so that our proof is complete.
If needed,  the
small bias introduced by a finite $N$ can be easily removed
just multiplying $\hat k_3,\hat \sigma$ derived from the likelihood method
by suitable functions of $N$.

The same calculation can be carried out in the more general case
of dipendent variables, but the search for the maximum is more
simply performed numerically when the situation is more complicated
(e.g., because of the presence of experimental
errors, or of more parameters, or more
complicate parameter dependence). We just quote the result
when only  an overall skewness parameter is required, as when
the 3-point correlation function is given by $k_{ijk}=k_3 s_{ijk}$,
and the tensor $s_{ijk}$ is known (see Section 5). The best estimate for
$k_3$ is then
\beq
\hat k_3=-6 {s^{ijk}h_{ijk}\over s^{ijk}s^{lmn}h_{ijklmn}}\,,
\eeq
which reduces to the expression above when $s_{ijk}=\delta_{ij}
\delta_{jk}$,
using the relation
\beq\label{sums}
\sum_{i,j} h_{iiijjj}=\sum h_{6i}+(\sum_i h_{3i} )^2-
\sum_i h_{3i}^2\,,
\eeq
and observing that $(\sum h_{3i})^2$ is of order $\hat k_3^2$, and
thus negligible.

Once we have the best estimators $\hat \alpha_i(\vx)$
of our parameters, we need to estimate
the confidence regions for that paramaters, i.e. the range of values in
which we expect to find our estimators to a certain probability, given that
the data  distribution is approximated by the MEE.
The problem consists in determining  the behavior of the
unknown distribution $P[\hat\alpha_i(\vx)]$, when we know the
distribution for the random variables $x_i$. This problem is
generally unsoluble analitycally, and the common approach is
to resort to MonteCarlo simulations of the data. However,
we can always approximate $P(\hat\alpha_i)$ {\it around its peak}
by a Gaussian distribution multivariate {\it in the parameter space};
if the number of data $N\to\infty$, this procedure can be justified
by the central limit theorem. For instance, if $\hat k_3=\sum x_i^3/N$,
then its distribution will tend to a Gaussian whatever the distribution
of the data $x_i$ is, in the limit of large $N$. In more general cases
(e.g. correlated data) the central limit theorem does not guarantee
the asymptotic Gaussianity; we can expect however it to be a first
reasonable approximation far from the tails.
 If this approximation is adopted, then
it can be  shown (see e.g. Kendall, Stuart \& Ord 1987) that
the covariance matrix of the parameters can be written as
\beq
\Sigma_{ab}^{-1}=-{\partial \log L(\vx,\alpha_a)\over
\partial \alpha_a\partial\alpha_b}\Big|_{\alpha_a=\hat\alpha_a}\,,
\eeq
where $a,b$ run over the dimensionality $P$ of the parameter space. The
1$\sigma$ confidence region is then enclosed inside the $P$-dimensional
ellipses with principal axis equal to  $\lam_a^{1/2}$, where
$\lam_a$ are the eigenvalues of $\Sigma_{ab}$.
Let us illustrate this in the same simplified case as above: $N$
independent data characterized by variance $\alpha_1=
\sigma$ and skewness $\alpha_2=k_3$.
To further simplify, we assume that the mixed components
$\Sigma_{12}=\Sigma_{21}$ can be neglected (see below).
The component $\Sigma_{22}$ is then easily calculated as
\beq
\Sigma_{22}^{-1}=-\sum_{i,j}h_{iiijjj}/36\,,
\eeq
Thus, using Eq. (\ref{sums}),
the variance of $\hat k_3$ turns out to be
(dropping the hats here and below)
\beq
\Sigma_{22}= 6\sigma^6/N\,,
\eeq
which, not unexpectedly, is the sample skewness variance, i.e.
the scatter in the skewness of Gaussian samples
(for the dimensionless
skewness  defined as $\gam_1=k_3/\sigma^3$
 the variance   is $6/N$).
 In other words,
to this order of approximation, the variance in the sample skewness in
non-G data equals the variance in the sample
skewness of Gaussian distributed data.
The generalization to dependent data is
\beq\label{vark3}
\Sigma_{22}^{-1}=-s^{ijk}s^{lmn}h_{ijklmn}/36
\eeq
which gives then the variance of the estimator $\hat k_3$ in
the general case.

 More interesting is the error in the variance parameter $\sigma$ when
not only a non-zero skewness $k_3$ is present, but also a non-zero
kurtosis parameter $k_4$, defined in a way similar to $k_3$ as
$k_{ijkl}=k_4 s_{ijkl}$. Then the result turns out to be, in the
same approximations as above,
\beq\label{varvar}
\Sigma_{11}={\sigma^2\over 2N}\left[
1+\gam_2/2\right]\,,
\eeq
where $\gam_2=k_4/ \sigma^4$ is the dimensionless kurtosis.
The mixed components amount to $\Sigma_{12}^{-1}=\Sigma_{21}^{-1}=
-Nk_3/\sigma^7$. Then we see that in the determinant
of $\Sigma_{ab}$ we have $[\Sigma_{12}^{-1}]^{2}\ll
[\Sigma_{11}^{-1}\Sigma_{22}^{-1}]$ for $k_3/\sigma^3\ll 1$, which
is again the mild non-Gaussianity condition we are assuming
throughout this work.

Eq. (\ref{varvar}) is again an expected results: it is infact the
variance of the standard deviation $\sigma$
for $N$ independent data when a non-zero fourth-order
moment is included.
The first term in (\ref{varvar})
is the usual variance of the sample variance for
Gaussian, independent data.
 The second term is due to the
kurtosis correction: it will broaden the CR for $\sigma$ when $k_4$
 is positive,
and will shrink it when it is negative.
Depending on the relative amplitude of the higher-order
corrections, the CR for the variance can extend or reduce.
It is important however to remark that this estimate of the
confidence regions is approximated, and that it can be trusted only
around the peak of the likelihood function. This means that we cannot
use the CR estimated by the method  exposed here when we are interested in
large deviations from the best estimates. The true CR will
in general be more and more different from this simple estimate as its
probability content grows: to a confidence level of, say, 99.7\%,
we cannot reliably associate a CR of 3$\Sigma_{11}^{1/2}$,
as we would do were the $P(\hat\alpha_i)$ a perfect Gaussian distribution.

This limitation is the main motivation for the rest of this work.
Adopting a $\chi^2$ technique, we will be able to give an
analytical expression for the CR of the variance (or of other
parameters entering $\chi^2$) when non-G corrections are
present. This is useful whenever
we actually measured non-vanishing higher-order cumulants and wish to
quote a CR for the variance allowing for the non-Gaussianity, or when,
more generally, we have reason to suspect that our data are non-Gaussian
and we wish to investigate how the CR vary with different non-Gaussian
assumptions. As already remarked, non-Gaussianity
can also be invoked to put in
agreement two  experimental
results reporting non-overlapping CR.
The results of the next sections
will confirm the approximate trend  of
Eq. (\ref{varvar}), as long as the CR
does not extend into the tails of the paramater distribution.

\section{Non-Gaussian $\chi^2$ method}

If our data are distributed following the MEE, then we can measure
the likelihood to have found our actual dataset integrating
the LF over all the possible outcomes of our experiment.
According to  the
$\chi^2$ method,
the actual dataset is more likely to have occurred, given
our assumptions, the
higher is the probability to obtain values of $\chi^2$ larger than the
measured $\chi^2_0$.
Then the relevant integral we have to deal with is
\beq\label{rel}
M(\chi_0)=\int_{\chi^2\le\chi^2_0} L(x,\lam) \prod_i dx_i\,,
\eeq
where the region of integration extends over all the possible
data values which lie inside the region delimited by
the actual  value $\chi^2_0$.
The function $M(\chi_0)$
gives then the probability of occurrence of
 a value of $\chi^2$ smaller
than the one actually measured.
We can then use $M(\chi_0)$ for evaluating a CR for the
parameters which enter $\chi_0^2$, like the quadrupole and the primordial
slope in the case of CMB. The CR will depend parametrically
on the higher-order moments; however, this will not provide a CR
for the higher-order moments themselves.  The method of the
previous section can always be employed to yield a first
approximation for such moments.
Both too high and  too small
values of $\chi^2_0$ have to be rejected; fixing a confidence level of
$1-\eps$, we will consider acceptables the values of the parameters for
which $M(\chi_0)$ is larger than $\eps/2$ and smaller than $1-\eps/2$.
Notice that the theoretical parameters enter $M(\chi_0)$ both
through the integrand $L(x,\lam)$ and  the integration region
$\chi^2_0$. This section is devoted to the evaluation of (\ref{rel}).

Let us split the LF into  a  Gaussian part and a
non-G correction,
\beq
L=L_g+L_{ng}\,.
\eeq
The integral of the Gaussian part is the standard one, and
it is easily done:
\beq
\int G(x,\lam) \prod_i dx_i=\int G(y,\delta) \prod_i dy_i
=\int_0^{\chi_0^2} P_N(\chi^2)d \chi^2=F_N(\cz)\,,
\eeq
where $P_N(\chi^2)$ is the $\chi^2$ PDF with $N$ degrees of freedom,
and $F_N$ its cumulative integral.
Notice that the integral has been performed over a compact
region in $x$-space whose boundary is $\chi^2=\chi_0^2$. It is
then convenient  to change variables
 in the integral, from $y^j$
to the hyperradius $\chi^2$ and
$N-1$ angles
\beq
\int \prod_{i}dx_i\to\half \int (\chi^2)^{p-1}d(\chi^2)
d\Omega={A_{2p}\over 2}\int (\chi^2)^{p-1}d(\chi^2)\,,
\eeq
where $p=N/2$ and
$A_{2p}=2\pi^p/\Gamma(p)$ (the surface area
of a  unitary $(N-1)$-sphere).
The same procedure will be applied several times in the following.

For the non-G sector we have to consider
separately the three last terms in the MEE (\ref{mee}).
However, since it will be shown
shortly that the term linear in the
skewness  does not contribute to the final
result, we will focus only on the two last terms in the MEE.
Let us consider the term in $k^{ijkl}$. Its integral can be
written as
\beqa
K&=&{ k^{ijkl}(x)\over 24 }\int G(x,\lam) h_{ijkl}(x,\lam) \prod_i dx_i
\nonumber\\
&=&
{ k^{ijkl}(y)\over 24 }\int G(y,\delta) h_{ijkl}(y,\delta) \prod_i dy_i=
{ k^{ijkl}(y)\over 24 }\int \partial_i\partial_j\partial_k\partial_l
\prod_i G_i dy_i\,,
\eeqa
where $G_i=(2\pi)^{-1/2} \exp(-y_i^2/2)$ and where the kurtosis
tensor $k^{ijkl}$ in the first line is to be
calculated with respect to $x^i$, while in the last
line with respect to $y^i$.
Since $y^i=A^i_a x^a$, then
\beq\label{tensor}
k^{ijkl}(y)=A^i_aA^j_bA^k_cA^l_d k^{abcd}(x)\,,
\eeq
and likewise for $k^{ijk}$.
Suppose now
one of the subscripts,
say $i$, appears an odd number of times,
 like in $k^{iiij}$. Let us call
then $i$ an odd index. The integral $K$ will then be
odd in $y_i$. Since the integration region is symmetric around the
origin, $K$ would vanish. This shows that any term in
$K$ containing odd indexes of $k^{ijkl}$ must vanish.
This explains also why  the skewness term $h_{ijk}$,
which  always contains some odd
index,  gives no contribution to the  likelihood
integral.
The only non-zero terms in $K$
are then of the type $k^{jjkk}$ and
$k^{jjjj}$ (the index order is irrelevant). We then need only
two kinds of integrals for as concerns $K$.
Let us evaluate the first kind:
\beq
I_1=\int \prod_{i\not= j,k} G_i dy_i[\int dy_j dy_k
\partial^2_j\partial^2_k
 G_j G_k]\,.
\eeq
The inner integral must be evaluated inside the circle bounded
by $\rho^2_{jk}=\chi_0^2-\chi_{jk}^2$, where $\chi^2_{jk}=
\sum_{(i\not= j,k)} y_i^2$. Transforming the variables $(y_j,y_k)$
to the radius $\rho_{jk}$
and the angle $\theta$, and integrating over the new variables, we obtain
\beq
I_1=\int (\prod_{i\not= j,k} G_i dy_i) G_2(\rho_{jk}) f_1(\rho_{jk})
=G_N(\chi_0) \int f_1(\rho_{jk}) \prod_{i\not= j,k} dy_i\,,
\eeq
where we define $G_N(\cz)=(2\pi)^{-N/2} \exp(-\cz^2/2)$, and where
\beq\label{f1}
f_1(\rho_{jk})=\pi(4\rho_{jk}^2-\rho_{jk}^4)/4\,.
\eeq
Changing again variables under the integral to the
hyperradius $\chi^2_{jk}$ and $N-3$ angles, we obtain
\beq
I_1=\half
G_N(\chi_0) A_{2p_2}\int_0^{\chi_0^2}
 f_1(\rho_{jk}) (\chi^2_{jk}) ^{p_2-1}d \chi^2_{jk}=q_1(\chi_0) G_N(\chi_0)\,,
\eeq
where  $p_2=(N-2)/2$, and where
\beq\label{q1}
q_1(\cz)= \half \pi^{N/2}\cz^N[N+2-\cz^2]/\Gamma(2+N/2)\,.
\eeq
All the other integrals we need can be obtained in similar ways.
For instance, the second kind of non-vanishing integral in $K$ is
\beq
I_2=\int \prod_{i\not= j} G_i dy_i[\int dy_j
\partial^4_j
 G_j ]=\half G_N(\chi_0) A_{2p_1}\int_0^{\chi_0^2}
 f_2(\rho_{j}) (\chi_j^2)^{p_1-1} d \chi^2_{j}=q_2(\cz)G_N(\cz)\,,
\eeq
where $p_1=(N-1)/2$ and
\beq\label{q2}
q_2(\cz)={3\over 2}\pi^{N/2}\cz^N\left[
N+2-\cz^2
\right]/\Gamma(2+N/2)\,.
\eeq
For as concerns the last term in (\ref{mee}), we need only to
evaluate three new integrals, from the terms
 $h_{jjkkll}$, $h_{jjjjkk}$ and $h_{j...j}$.
Let us denote these integrals by $I_3, I_4$ and $I_5$. In
 complete analogy to the two integrals $I_1,I_2$, we find
$I_i=G_N(\cz)q_i(\cz)$ where
\beqa
q_3(\cz)&=&  {1\over 4}\pi^{N/2}
\cz^N h(\cz)\,,
\nonumber\\
q_4(\cz)&=&
{3\over 4}\pi^{N/2}\cz^N h(\cz)\,,
\nonumber\\
q_5(\cz)&=&{15\over 4} \pi^{N/2}\cz^N h(\cz)
\,,
\eeqa
and where
\beq
h(\cz)=
\left[-{4\over \Gamma\left(1+N/2\right)}
+{4\cz^2\over \Gamma\left(2+N/2\right)}
-{\cz^4\over \Gamma\left(3+N/2\right)}\right]\,.
\eeq

These expressions are all what we need for the complete
evaluation of the likelihood integral.
The general result is then
\beqa\label{finres}
&&M(\cz)=\int L \prod dx_i=F_N(\cz)\nonumber\\
&&+{G_N(\chi_0)\pi^{N/2}\cz^N \over 2\Gamma(2+N/2)}
\left[ C_a\left(N+2-\chi_0^2\right)+C_b
\left(-N-2+2\chi_0^2-{\chi_0^4\over N+4}\right)\right]\,,
\eeqa
where
$C_a=c_1+3 c_2$, and $C_b=c_3+3 c_4+15 c_5$, and
the coefficients $c_i$ are formed by summing over all
the even diagonals of the correlation tensors $k^{ij..}$ and multiplying
for the Edgeworth coefficients $(1/24)$ for $c_1,c_2$ and $(1/72)$
for $c_3,c_4$ and $c_5$. Let us denote with
 ${\rm Tr}_{ab..}(k^{ij...})$ the
sum over all the disjoint partitions of $a$-plets, $b$-plets, etc.
of equal indexes: e.g., ${\rm Tr}_{22}(k^{ijkl})$ means summing over all
terms like $k^{iikk},k^{ikik},k^{ikki}$ but without including
$k^{iiii}$; or, ${\rm Tr}_{24}(k^{ijk}k^{lmn})$ means summing
over terms like $k^{iii}k^{ijj}$ or $k^{iij}k^{jii}$.
 Then, the
 coefficients $c_i$ in Eq. (\ref{finres}) can be written as
\beqa\label{fincoeff}
c_1&=&(1/24){\rm Tr}_{22}(k^{ijkl})\,,\qquad
c_2=(1/24){\rm Tr}_{4}(k^{ijkl})\,,\nonumber\\
c_3&=&(1/72){\rm Tr}_{222}(k^{ijk}k^{lmn})\,,\quad
c_4=(1/72){\rm Tr}_{24}(k^{ijk}k^{lmn})\,,\nonumber\\
c_5&=&(1/72){\rm Tr}_{6}(k^{ijk}k^{lmn})\,.
\eeqa
Let us make some comments on the result so far obtained.
First, notice that $M(\cz)$ is a cumulative function and as such
it has to be  a monotonically increasing function of its argument
bounded by zero and unity. This provides a simple way to check
the consistency of our assumptions: when the higher-order moments
are too large, the MEE breaks down,  $M(\cz)$ is no longer
monotonic, and can decrease below zero or above unity.
Second, let us suppose that
the higher-order correlation
functions are positive, which is the case for the galaxy
clustering (see  Section 6).
Then the non-G corrections in Eq. (\ref{finres})
are negative for  $\cz^2\gg N$.
The value of $\cz(\eps)$, corresponding to a  probability content
$M(\cz)=1-\eps$,
is a measure of how large is the confidence region
associated with the threshold $\eps$, if $\cz$ is single-valued on
 the parameter space, a common occurrence in practice.
The fact that the corrections
are negative for $\cz^2\gg N$  implies that the value of
$\cz=\cz(\eps)$   is larger than in the purely
Gaussian case, in the limit of $\eps\to 0$.
 Consequently, if the higher-order
correlation functions are positive,
{\it the confidence regions are
systematically widened when the non-Gaussian corrections
are taken into account}.  For $\eps$ not very close to zero
is not possible to make such
a definite statement; the regions of confidence will
widen or narrow depending on the value of the moments, as will
be graphically shown in the next section.
Let us remember that   for two-tail tests the CR is
enclosed by  the upper limit
$\cz^{(1)}(\eps/2)$ and the lower limit
$\cz^{(2)}(1-\eps/2)$, and the limits behave differently
depending on $\eps$ and on the higher-order moments.
Finally, it is easy to write down
the result in the particular case in which all the cumulant
matrices are diagonal, i.e. for statistically independent variables.
In this case the variables $y^i$ are simply equal to $x^i/\sigma_i$, if
$\sigma_i=(\lam^i_i)^{-1/2}$, and we can put
 $k^{iii}(y)=k^{iii}(x)/\sigma^3\equiv\gam_{1,i}$, and likewise
 $k^{iiii}(y)\equiv \gam_{2,i}$
(skewness and kurtosis coefficients). Then, we have $c_1=c_3=c_4=0$, and
Eq. (\ref{finres}) can be simplified to
\beq\label{resind}
M(\cz)=F_N(\cz)+G_N(\chi_0) q(\cz)\,,
\eeq
where
\beq\label{indp}
q(\cz)={6\pi^{N/2}\chi_0^N \over (N+2)\Gamma(N/2)}
\left\{
{\gam_2\over 24}\left[(N+2)-\chi_0^2\right]+{5\over 72}\gam_1^2
\left[-(N+2)+2\chi_0^2-{\chi_0^4\over N+4}\right]\right\}\,,
\eeq
and where we introduced the average squared skewness, $\gam_1^2=
\sum \gam_{1,i}^2/ N,$ and the average kurtosis,
$\gam_2=\sum \gam_{2,i}/N$.

\section{Graphical examples}

This section is devoted
to illustrate graphically some properties of the
 function $M(\cz)$ in its
simplified version (\ref{resind}) above,
first putting $\gam_1=0$, then $\gam_2=0$, and assuming
 $N=10$ and $N=100$.
In all this section we can think of
 $\cz$ as depending monotonically on
one single parameter, for instance the overall normalization $A>0$
of the correlation function:
$\cz^2(A)=x^i x^j (A c_{ij}+e_{ij})^{-1}$.
Then it appears that $\cz^2$ decreases from a finite, positive value
to zero as $A$ goes from zero to infinity. The range in which
$A$ is bounded increases or decreases with the bounding range of $\cz$;
we can then speak of a CR on $\cz$ meaning in fact the corresponding
CR on the parameter $A$. In the general case, the relation
between $\cz$ and its parameters can be quite more complicated.
In Fig. 1{\it a} (for $\gamma_1=0$ and $N=10$),
 we show
how the function $M(\cz)$ varies with respect to the non-Gaussian
parameter $\gamma_2$.
Schematically, for $\cz^2/N> 1$, the function $M(\cz)$ decreases
when $\gamma_2>0$ and increases in the opposite case.
As anticipated,
for too large a $\gam_2$, $M(\cz)$ develops a non-monotonic behavior.
The consequence of the behavior of $M(\cz)$
 on the confidence region of
$\cz$ is represented in Fig. 1{\it b}, where the contour plots of the
surface $M(\cz,\gamma_2)$ are shown.
Consider for instance the two outer contours, corresponding to
$M=.01$, the leftmost, and $M=.99$, the rightmost.
The important point is that
the range of $\cz$ inside
such confidence levels increases for increasing $\gamma_2$;
with respect to the Gaussian case, $\gamma_2=0$, the acceptable
region for $\cz$ widens substantially
even for a  small non-Gaussianity.
As a consequence, a value as high as, say, $\cz^2/N=2.5$, is
inside the 99\% confidence level if $\gam_2>.2$.
The same is true for the other contour levels, although
with a  less remarkable trend. This behavior confirms the approximate result
of Eq. (\ref{varvar}).
As anticipated,  this means
that the non-G confidence regions  will be larger and larger (if
the higher moments are positive) than
the corresponding Gaussian regions for higher and higher
probability thresholds. Notice that in this case the parameter
$\gam_2$ itself cannot be given a CR, since as we already noticed
the LF has no maximum when
varied with respect to it. We can use $\gam_2$ only as an external
parameter, either provided by theory, or
estimated from the data in some other way.
Fig. 2{\it a,b} reports the same features
for $\gamma_1=0, N=100$. Now the confidence regions are much narrower,
because of the increased number of experimental data.

The situation is qualitatively different considering $\gamma_2=0$
and varying $\gam_1$, the average skewness.  Fig. 3 and Fig. 4
are the  plots for this case ($N=10$ and $N=100$,
respectively).
The contour levels are obviously symmetric for $\pm \gam_1$.
Now for any given $\cz$ there is a CR for $\gam_1$ and
viceversa, so that a bound can be given on each parameter
given the other one,
although the joint CR for {\it both} parameters is infinite.
Now one can see two
different features in the contour plots.
For the outer contours,
delimiting levels of 1\% on both tails, the CR of $\cz$ {\it increases}
for larger $|\gam_1|$,
with a minimum for the Gaussian case.
For the internal contours, however, the CR actually shrinks for
larger $|\gam_1|$,   being maximal at the Gaussian point.
It is clear that in the general case, $\gam_1,\gam_2\not=0$,
the topography of the LF can be quite complicated.

\section{Comments on practical application}

The results of the previous sections can be employed to
estimate theoretical parameters and confidence regions in
several interesting cases. We consider here two of these,
the large scale structure (LSS) of galaxies
and the CMB.

In the case of LSS surveys, the data usually consist of the
fluctuations $x^i=\delta n^i/\hat n$ in the number counts of galaxies in
the $i$-th cell in which the survey is partitioned.
(We assume here for simplicity that the average density $\hat n$ is fixed
{\it a priori}. Otherwise, we can include it in the set of
parameters to be estimated.)
The main problem in applying the formalism developed so far
to real situations is  to choose a ``good'' set of theoretical
parameters $\alpha_j$. In principle we can parametrize the
statistical properties of the LSS in an infinite number of ways.
However, the particular set of parameters we are going to adopt
has been singled out
in the current literature, both theoretical and observational,
with very few exceptions.
Assuming for the correlation function the power-law form
$\xi(r)=(r_0/r)^{\gam}$,  the cell-averaged
$c^{ij}$ is given by the following
expression
\beq\label{cfint}
c^{ij}=\int \xi(r_{12})W_{R_i}(\vr_1)W_{R_j}(\vr_2) d\vr_1 d\vr_2\,,
\eeq
where $r_{12}=|\vr_1-\vr_2|$, and $W_{R_i}$ ($W_{R_j}$)
 is the normalized window function
of characteristic size $R_i$ ($R_j$)
relative to the $i$-th ($j$-th) cell.
If the cells $i,j$ are fully characterized by a size $R$ and a separation
$s_{ij}$, the integral (\ref{cfint}) can be written as
\beq
c^{ij}=J(\gam,R/s_{ij})(R/r_0)^{-\gam}\,,
\eeq
where $J(\gam,R/s_{ij})$ is a dimensionless function of $\gam$ and $R/s_{ij}$.
Following standard work (e.g. Peebles 1980)
we will then
write for the higher-order correlation functions the following expressions
% (no repeated index summation)
\beqa\label{lss-cf}
k^{ijk}&=&Q(c^{ij}c^{jk}+c^{ik}c^{jk}+
c^{ik}c^{ij})\,,\nonumber\\
k^{ijkl}&=&R_a\sum_2 c^{ij}c^{jk}c^{kl}+R_b\sum_3c^{ij}c^{ik}c^{il}\,,
\eeqa
where $\sum_2$ ($\sum_3$) means summing over all the 12 (4)
tree graphs with at most
two (three) connecting lines
per vertex (i.e. summing over topologically equivalent
graph configurations).
Note that we define $Q,R_a$ and $R_b$ in terms of the {\it cell-averaged}
correlation functions, rather than in terms of $\xi(r)$, as currently done.
Our definition has the advantage that
from $Q,R_a$ and $R_b$ one can obtain directly the often quoted
scaling coefficients $S_3=3Q$ and $S_4=12R_a+4R_b$, without complicated
integrals over the window functions. The drawback is that our $Q,R_a,R_b$
cannot be compared directly to the values reported in literature, albeit the
difference should be very small.

Several analysis of large scale surveys show that $Q,R_a,R_b$ are fairly
constant over several scales, and of order unity. On scales larger that
$\approx 10\hm$, however, the power-law
form of $c^{ij}$ is not longer acceptable. For such scales
is preferable to parametrize instead the power spectrum
$P(k)$ and to use the identity
\beq\label{corrpk}
\xi(r_{ij})={1\over 2\pi^2r_{ij}}\int_0^{\infty} k P(k) \sin(kr_{ij})
  dk\,,
\eeq
from which, using Eq. (\ref{cfint}),
\beq
c^{ij}={1\over 2\pi^2s_{ij}}\int_0^{\infty} k P(k) W_k^2\sin(ks_{ij})
  dk\,,
\eeq
where $W_k$ is the Fourier transforms of the window function.
Various forms of $P(k)$ have been proposed so far.
For instance, one can assume the simple
functional form proposed by Peacock (1991), with its two
scale parameters $k_0,k_1$, or the CDM-like form of
Efstathiou, Bond \& White (1992), involving
an overall normalization and a dimensionless parameter $\Gamma$.
To give an idea of how big the
non-G corrections are, let us assume to have $N$ independent data
(i.e., data on cells at separations much larger than the correlation length)
and let us  estimate the parameters $\gamma_1,\gamma_2$ of
Eq.  (\ref{indp}). Since $\gam_{1,i},\gam_{2,i}$ and $\sigma_i$
are the same for all the
$N$ data, one has (dropping the subscript $i$)
   $\gam_2=k^{iiii}(x)/ \sigma^4=S_4\sigma^2$ and
$\gamma_1^2=S_3^2\sigma^2$. For $S_3\approx 3$ and $S_4\approx 20$,
as large scale surveys suggest, one gets $\gamma_2\approx 20\sigma^2$, and
$\gamma_1^2\approx 10
\sigma^2$. For scales around 10 $\hm$ or so, where
$\sigma^2\approx 1$, $\gamma_1,\gamma_2$ are then
very large, but they decrease rapidly for larger scales. On scales
larger than
30 $\hm$ or so, $\gamma_1,
\gamma_2$ are small enough to use the MEE also near the tails.

The non-G
LF allows a determination of the parametric set
in   such a way that the best estimate of one parameter depends on all
the other ones, unlike the common procedure
of estimating one parameter fixing the others  (in particular,
fixing the non-Gaussian parameters to zero).
For instance, the usual way of estimating $r_0,\gam$ is to find the
best $\chi^2$ power-law fit to the observed correlation function,
which amounts to assume a Gaussian distribution around the mean values.
Both the estimate and
the confidence region would  then be corrected by the higher order terms.
However, as already mentioned, we can use the MEE for
estimating the higher-order moments themselves only if enough terms
have been included in the expansion.
The reason is clear by looking at the Eq. (\ref{mee}): at this order
of truncation,
the expansion is linear in the fourth order moment, and as
a consequence it has no maximum when derivated with respect to,
e.g., $R_a$ or $R_b$. The best estimate does not exist at all.
We can  give however an estimate for $Q$, and we can expect it to
be a good estimate as long as it is
in the regime in which the MEE holds.
 A simple way to check this is
to see whether for that value of $Q$ the function $M(\cz)$ is
well-behaved, i.e. is a monotonic increasing function bounded
by zero and unity. In principle, one
can proceed further, including more and more terms in the LF,
so that one can reach  not only  a higher degree of approximation,
but also estimate the error introduced by the truncation itself.
Needless to say, these goods come
at the
price of a factorial increase in algebraic complication.

Once we have chosen our
parameter set, the only remaining difficulty is to evaluate  the coefficients
$c_1,..c_5$.
Let us remark that the Eqs. (\ref{lss-cf}) are valid with respect
to the original
data $x^i$, while we need the correlation functions for
$y^i=A^i_j x^j$ to evaluate $c_1,..c_5$. The relation between the two sets
of correlation functions is provided by Eq. (\ref{tensor}).
The evaluation of $c_1,..c_5$ is  straightforward.
One needs simply to scan all the possible combinations of indexes
$i,j,k,..$ and sum only those tensor components with
all equal indexes (for $c_2$ and $c_5$), or those with all
paired indexes (for $c_1$ and $c_3$), or finally those with a $(2,4)$
index structure (for $c_4$).  A more explicit
expression for the coefficients can also be found in specific
cases (e.g. exploiting
the symmetry under index permutation of the tensors in  (\ref{lss-cf})),
 but the general calculation can be coded so easily on computers
that we prefer to leave it in the form (\ref{fincoeff}).
%\def\cij{c_{ij}}
%\beqa\label{set}
%c_1&=& \half \sum_{i\not=j}\left[ R_a c_{ij}(c_{ij}^2+c_0^2+\cij c_0)
%+R_b\cij^2 c_0\right]\\
%c_2&=& {3R_a+R_b\over 6}N c_0^3\\
%c_3&=& {Q^2\over 8}\sum_{i\not=j\not=k}
%\left[ c_{ij}c_{jk}(2c_0+\cij)(2c_0+c_{jk})+4(c_{ij}c_{jk}+c_{ik}c_{jk}+
%c_{ik}c_{ij})^2\right]\\
%c_4&=& {Q^2\over 8} \sum_{i\not=j} c_{ij}(2c_0+c_{ij})
%[2c_0^2+2c_0\cij+\cij^2]\\
%c_5&=& {Q^2\over 8}N c_0^4
%\eeqa
Let us then summarize the steps needed to analyze a given set of data.
First, one selects  a value for the chosen parameter set inside a
plausible range. Second,
one diagonalizes, for that particular
parameter set, the quadratic form $x^i\lam_{ij}x^j$
so to determine the matrix $A^i_j$ such that $y^i=A^i_j x^j$.
Third, one evaluates the five coefficients $c_1,..c_5$ summing
over all the required tensor components. Fourth,
one
evaluates $L$ and $M(\cz)$
for the selected parameter set. Fifth, one repeates the four previous steps
spanning a reasonable range in the parameter space. Finally,
the values
for which $L$ has a maximum inside the range, if any, are the best
estimate of the set of parameters, while the region for which $
\eps/2< M(\cz)<1-\eps/2$
defines their joint confidence region.

For the CMB, the procedure is very similar.
 The major
difference is the set of parameters we are interested in.
For simplicity, let us consider an experiment like COBE, in which the
large angular beam size is mainly designed to study the Sachs-Wolfe effect
of primordial fluctuations.
The two-point angular correlation function can be
conveniently written as
\beq\label{cfa}
c^{ij}= \sum_{l=2}^{\infty} C_l W_l(\beta)P_l(\cos\alpha_{ij})\,,
\eeq
where $\alpha_{ij}$ is the angular separation between the $i$-th
and $j$-th pixel on the sky,
$W_l(\beta)$ is the observational window function relative to a
beam angular size $\beta$,
 $P_l$ is
the Legendre polynomial of order $l$, and $C_l$ is defined in terms
of the multipole coefficients $a^m_l$ as
\beq
C_l=\sum_{m=-l}^{l} |a^m_l|^2\,.
\eeq
For the Sachs-Wolfe effect of fluctuations with power spectrum
$P=Ak^n$ we can derive the expected variance of the amplitudes $a^m_l$ as
(e.g. Kolb \& Turner 1989)
\def\qps{Q^{PS}_{rms}}
\beq\label{sw}
\sigma_l^2\equiv <|a^m_l|^2>= {(\qps)^2 \over 5}{\Gamma[(9-n)/2]
\Gamma[l+(n-1)/2]\over
\Gamma[(3+n)/2]\Gamma[l+(5-n)/2]}\,,
\eeq
where $\qps$ is the expected quadrupole signal  derived from
the correlation function. The theoretical value for $C_l$ is then
$C_l=(2l+1)\sigma_l^2$, and it depends uniquely on $\qps$ and $n$.
Finally,  we rewrite Eq. (\ref{cfa}) as
\beq\label{cfa2}
c^{ij}= \sum_{l=2}^{\infty}(2l+1)\sigma_l^2 W_l(\beta) P_l(\cos\alpha_{ij})\,.
\eeq
The correlation function for the Sachs-Wolfe temperature
fluctuations is then parametrized by $\qps$ and $n$. The  situation for
the  higher-order correlation functions
is much less well established. Non-Gaussianity in the CMB is
predicted by several models, like topological defect theories, or
non-standard inflation, or can be induced
by some kind of foreground contamination.
 There is not, however, a single, widely accepted way
to parametrize non-Gaussianity in this context (see e.g.
Luo \& Schramm 1993 for some possible alternatives). A very simple
possibility is to assume for the CMB $n$-point
correlation functions the same kind
of scaling observed in LSS.
Preliminary constraints on the 3-point  parameter from the
COBE data have already been
published (Hinshaw \etal 1994).  Some model of inflation
predicts indeed this sort of scaling, although the expected amplitude
of the non-Gaussian signal in standard models
is far below observability
(Falk, Rangarajan \& Srednicki 1993;
Gangui \etal 1994).
For small scale experiments, the $c^{ij}$ parametrization
is different, and often
a Gaussian shape $c^{ij}=c_0\exp(-\alpha_{ij}^2/2\alpha_c^2)$,
is assumed.
The formalism here presented can be clearly
applied to any desired form of the correlation function.

\section{Conclusions}

Let us  summarize the results reported here.
This work is aimed at presenting a new analytic formalism for parametric
estimation with the maximum likelihood method for non-Gaussian
random fields. The method can be applied to a large class
of astrophysical problems.
The non-Gaussian likelihood function allows the determination
of a full set of parameters and their {\it joint} confidence region,
without arbitrarily fixing some of them,
 as long as enough non-linear terms are
included in the expansion.
The CR for all the relevant parameters can be estimated
by approximating the distribution function for the parameter
estimators around its peak by a Gaussian, as in Sect. 3.
 To overcome this
level of approximation, in Sect. 4
we  generalized the $\chi^2$ method to
include non-Gaussian corrections.
The most interesting result is then
that the CR
for the parameters which enter $\cz^2$ is systematically
widened by the inclusion
of the non-Gaussian terms, in the limit of $\eps\to 0$. Two
experiments producing incompatible results can then be brought
to agreement when third and fourth-order cumulants are introduced.
In the more general case, the CR may extend or reduce.

While we leave the analysis of real data
to subsequent work, we displayed some preliminary comments
on the application to two
important cases, large scale structure and cosmic microwave background.

There are two main limitations to the method. One is that we
obviously have to truncate the MEE to some  order, and consequently
 the data analysis  implicitly assumes  that all
the  higher moments   vanish.
The second limitation  is that the method is not applicable to strongly
non-Gaussian field, where the MEE breaks down.
This can be seen directly from Eq. (\ref{finres}): for arbitrarily
large constants $c_1-c_5$ the likelihood integral is not positive-definite,
although always converge to unity.
Assuming the scaling relation of Eq. (\ref{lss-cf}), for instance,
the condition $c^{ij}<1$
will ensure that the higher order terms are not dominating over
the lower terms,  as long as the scaling constants are of
order unity. Basing upon the current understanding of the matter
clustering, we expect the condition of weak
non-Gaussianity to hold for scales ranging from $\sim
30\hm$ to the horizon scale.

\vspace{.3in}
\centerline{\bf Acknowledgments}
It is very likely  that the completion of this work
would have been much more difficult, if possible at all,
without the crucial suggestions of
  Albert Stebbins, and the
useful discussions with  St\'ephane Colombi and Scott
Dodelson. I thank them all.
Further, I  thank the generous hospitality at
the NASA/Fermilab Astrophysics Center,
where most of this work has been done.

\newpage
\vspace{.5in}
\centerline{\bf References}
\vspace{.2in}

\noindent
Amendola L. 1994 \rapj, 430, L9\\
% Feldman H., Kaiser N., \& Peacock J., 1994, \rapj in press.\\
Chambers J. 1967, Biometrika 54, 367\\
Efstathiou G., Bond J.R. \& White S.D.M. 1992 \rmnras, 258, 1{\tiny P}\\
Falk T., Rangarajan R., \& Srednicki M. 1993 \rapj, 403, L1\\
Gangui A., Lucchin F., Matarrese S. \& Mollerach S. 1994,
\rapj in press\\
Hinshaw G. \etal \rapj, 431, 1\\
Juszkiewicz R., Weinberg D. H., Amsterdamski P.,
Chodorowski M. \& Bouchet F. 1993,
 preprint (IANSS-AST 93/50)\\
Kendall M.,  Stuart, A., \& Ord J. K., 1987, Kendall's Advanced
Theory of Statistics, (Oxford University Press, New York)\\
 Kofman L. \& Bernardeau, F. 1994, preprint IFA-94-19\\
Kolb E. W. \& Turner M. 1989,  The Early Universe (Addison Wesley, Reading)\\
Luo X. \& Schramm D.N. 1993 \rprl, 71, 1124\\
Matsubara T. 1994, preprint UTAP-183/94\\
McCullagh P. 1984, Biometrika, 71, 461\\
Peacock J.A., 1991, \rmnras, 253, 1{\tiny P}\\
Peebles P.J.E., 1981, The Large Scale Structure of the Universe (Princeton
University Press, Princeton)\\

\newpage
\centerline{\normalsize\bf Figure caption}
\vspace{.3in}
{\bf Fig. 1}
{\it a)} Plot of $M(\cz)$ as a function of $\cz^2/N$ and of the
dimensionless kurtosis $\gam_2$, for $\gam_1=0, N=10$.
 For $\gam_2=0$ we return to the usual $\chi^2$
cumulative function. Notice how for large kurtosis $\gam_2$ the
cumulative function $M(\cz)$ develops minima and maxima, indicating
that the MEE is breaking down.
{\it b)} Contour levels of $M(\cz)$ corresponding to
$M=.01,.1,.2,.3,.7,.8,.9,.01$, from left to right. Notice how
the limits for $\cz$ broaden for increasing $\gam_2$.
\vspace{.3in}

{\bf Fig. 2}
{\it a)}
Same as in Fig. 1{\it a}, now with more data, $N=100$.
{\it b)} Contour levels of $M(\cz)$ for the same values as
 in Fig. 1{\it b}. The
CR is now much smaller than previously.
\vspace{.3in}

{\bf Fig. 3}
{\it a)}
Same as in Fig. 1{\it a}, now with  $\gam_2=0$, $N=10$,
and varying $\gam_1$.
{\it b)}  Contour levels of $M(\cz)$ for the same values as
 in Fig. 1{\it b}.

\vspace{.3in}
{\bf Fig. 4}
{\it a)}
Same as in Fig. 1{\it a}, now with  $\gam_2=0, ~N=100$,
and varying $\gam_1$.
{\it b)}  Contour levels of $M(\cz)$ for the same values as
 in Fig. 1{\it b}.

\end{document}